# The Effect of Gravity Between Particles on the Shape and Resonant Structure of Planetary Rings

A.E. Rosaev

(FGUP NPC "NEDRA" Yaroslavl, Russia)

Planetary rings are a very complex system. It is a mixture of different-size collided particles. The electromagnetic forces can play an important role.

However, in spite of very small mass of particles, gravitational forces between particles always take place and cannot be neglected. In some cases, effect of mutual ring's particles interaction may be significant. On the other hand, by what way gravitation may be taken into account? It is natural, to consider motion of the test particle in the gravitational field of uniform ring. Perturbation force in this case calculated as an integral. In the limit, this integral can be substituted by sum with very large N. At the symmetric circular ring we can apply central configuration model. As it is well known, some of planetary rings are elliptic. It is interesting to generalize this model to a case ring with small eccentricity (quasi-central configurations).

In result, the stationary distribution of particles along elliptic ring is obtained. The gravitation interaction between particles in rings can change shape of non-circular rings, so it is differ from keplerian ellipse. The model of quasi-central configurations gives a way of estimation of this effect in dependence of ring particles properties.

The possible another effects of ring's particles gravity interaction are discussed. The resonance perturbation of the particle of planetary ring by distant satellite is considered in the restricted 3-body problem. After that, the gravitation interaction between ring's particles is taken into account. The shift between simple mean motion resonances and parametric resonance zones is detected. This shift depends from ring properties. Unexpectedly, the resonance's structure depends only from particle's density and number of particles in ring, but not depends from particles (or ring) mass. The results are applied to the system of Saturn rings. By varying ring's particles properties, it is possible to explain the observed shift between Cassini (and Encke) division position and Mimas exact commensurability. Observed surface density and thickness of rings restrict lower size of ring particles, which have main contribution into gravitation interaction, by half-centimeter size.

INTRODUCTION. MAIN EQUATIONS

Planetary rings are a very complex system. It is a mixture of different-size collided particles. The electromagnetic forces can play an important role. As it is well known, some of planetary rings are elliptic. However, in spite of very small mass of particles, gravitational forces between particles always take place and cannot be neglected. The main target of this paper is to show, that effect of mutual ring's particles interaction maybe significant. On the other hand, by what way to gravitation may be taken into account? It is natural, to consider motion of the test particle in the gravitational field of uniform (homogeneous) ring. Perturbation force in this case calculated as an integral:

$$S_0 = \int_0^{2\pi} \frac{m d\lambda}{(R_0^2 + R^2 - 2RR_0 \cos(\lambda))}$$

In the limit, this integral can be substituted by sum with very large N. At the symmetric circular ring we can apply central configuration model. In case of low-collisional ring with large particles, this discrete model has some evident advantages over continues model. Meyer K.R and Schmidt D.S. (Meyer., Schmidt, 1993) successfully apply it for description some phenomena inside Saturn rings. It is interesting to estimate the change of shape of the eccentric ring under gravitational interaction between particles, in dependence of particles properties.

The main forces between particles, in according with (Rosaev, 2001), where stationary distribution of ring particles is obtained:

$$F_r = -\frac{GM}{R^2} - \sum_{j=1}^{N-1} Gm_j \frac{2R_0 \sin^2(\alpha_j/2) + x}{\left(x^2 + (2R_0 \sin(\alpha_j/2))^2 (1+x/R_0)\right)^{3/2}}$$

$$F_{tg} = -\sum_{j=1}^{N-1} Gm_j \frac{2R_0 \sin(\alpha_j/2)\cos(\alpha_j/2)}{\left(x^2 + (2R_0 \sin(\alpha_j/2))^2 (1+x/R_0)\right)^{3/2}}$$

(1)

$$F_z = 0$$

where $R = R_0 + x$ - the distance of testing particle from the centre, $R_0$ - the distance of $j$-particle from the centre, $\Omega$ - angular velocity, $\alpha_j = 2\pi\, j/N + \varphi$, where $\varphi$ is possible angular declination from stationary position.

The problem of shift of the gaps in Saturn rings is well known and not completely be understand for present. However, it may be explained by taking into account gravity between particles and parametric resonance model.

First of all, review some important observation data, related with rings dynamics at resonance. The stellar occultation measurements by the photopolarimeter provided a typical radial resolution of $\sim$< 300 m. The edges of the rings at the gaps that do exist are so sharp, however, that the rings must be $\sim$< 200 m thick at those locations (Stone *et al.*, 1982.). **It means, that planar model of rings is possible.**

Clumps of material in the F ring were observed over 15 orbits (Stone *et al.*, 1982.). **It means, that structures like as central configurations really present at planetary rings, and moreover, maybe collisionless at finite time interval.** The model of the central configurations successfully applied to the real planetary rings by Meyer K.R. and Schmidt D.S. (Meyer, Schmidt, 1993).

The outer edge of the B-ring is located within the 2:1 mean motion resonance with the satellite Mimas. A detailed kinematic fit of the edge (Porco et al. 1984) shows that the ring edge's pattern speed is consistent with Mimas' mean motion. However, the edge does not fall exactly at the location of the resonance. **Instead, the edge semi-major axis is offset by 24 km outside the resonance location** (Namouni, Porco, 2002).

The extremely sharp boundary between the B and C rings does not show any gap, nor does it correspond in location to any significant resonance (Smith *et al.* 1981).

Although the Mimas 5:3 resonance is nearby, the Encke division, at a radial distance of 133,200 km corresponds better with one of five other features of great interest in the outer A ring (Smith *et al.* 1981).

It is confirm the main conclusion: **The gaps at Saturnian rings are offset by few km outside the resonance location.**

# THE STATIONARY DISTRIBUTION OF PARTICLES AT ELLIPTIC RING

Put ring of N particles m with small eccentricity in gravity field of central mass M. It is naturally to find stationary distribution of particles by true anomaly and about influence of mutual perturbations of particles on ring's eccentricity.

There are two possibility. For the low-massive particles m<<M, their orbital distribution determined by keplerian angular velocity (L - angular momentum):

$$\delta \approx const \frac{d\nu}{dt} = \frac{L}{mr^2} \approx \delta_0 (1 + e\cos(\nu)) \qquad (2)$$

The distribution of low-massive particles is shown at the Fig.1.

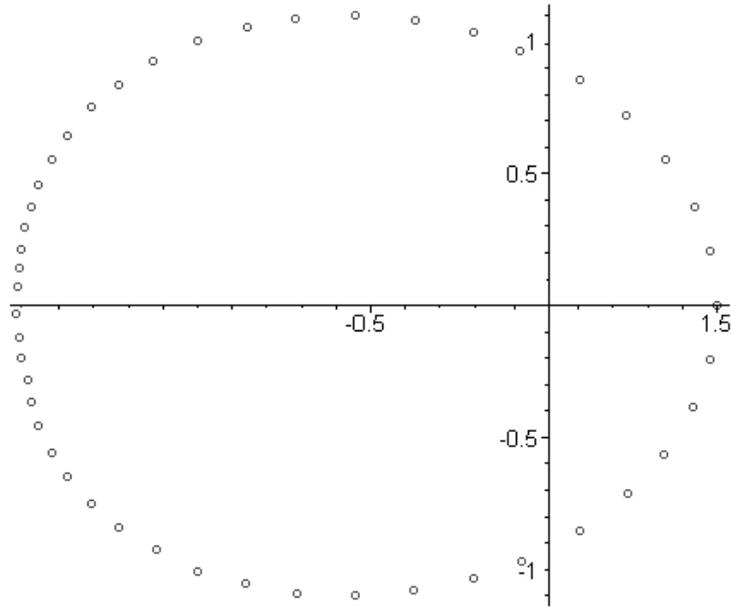

Fig.1. The orbital distribution of particles in eccentric ring (small mass particle case)

In the second case large particles the increasing of particles concentration near apocenter is expected. It is important, however, that concentration near apocenter take place in both cases large and small particles.

## THE SHAPE OF ECCENTRIC RING

The radial perturbing force become function of true anomaly ν:

$$F_r = -\sum Gm_i / \left|(2R(\nu))^2 \sin\alpha_i(\nu)/2\right| \approx -\frac{Gm}{2a^2(1-e^2)^2} \frac{(1+e\cos\nu)^2}{|\sin(\delta/2)|} \qquad (3)$$

where δ is determined by (2). The equations in orbital elements:

$$\frac{dp}{dt} = 2rF_{tg}$$

$$\frac{de}{dt} = \sin\nu\ F_r + (\cos\nu + (\cos\nu + e)r/p)F_{tg}$$

$$\frac{d\omega}{dt} = \frac{-\cos\nu}{e}F_r + \frac{\sin\nu}{e}(1 + p/r)F_{tg}$$

$$r = p/(1 + e\cos\nu)$$

(4)

Here p, e, ω - orbital elements (orbital parameter, eccentricity, perihelion argument), r – heliocentric distance, ν - true anomaly. The elements of orbits at the right sides are constant at integration.

An averaged tangential force in all cases (approximately) is equal zero. Radial force depends from particles distributions by longitude, which can vary from uniform to stationary. It always changes shape of ring. Really, eccentricity at pericenter differ form it at apocenter. From equation for eccentricity:

$$e = e + \Delta e(\nu);\quad \Delta e(\nu) \approx -\frac{Gm}{2a^2(1-e^2)^2}\frac{(1+e\cos\nu)^2}{|\sin(\delta_0(1+e\cos(\nu)/2)|}\sin\nu$$

(5)

It means, that eccentricity of ring in this model depends from true anomaly (Fig.2, 3). This result may be explains of the complex shape of some planetary rings. Note, this effect always takes place, independently from collisions, viscosity, electromagnetic forces, etc. At different conditions, the value of them is differ, but in some cases it may has dominant role and to be observable.

In addition, gravitational interaction between particles occur precession of orbit (equation for ω); it is well known.

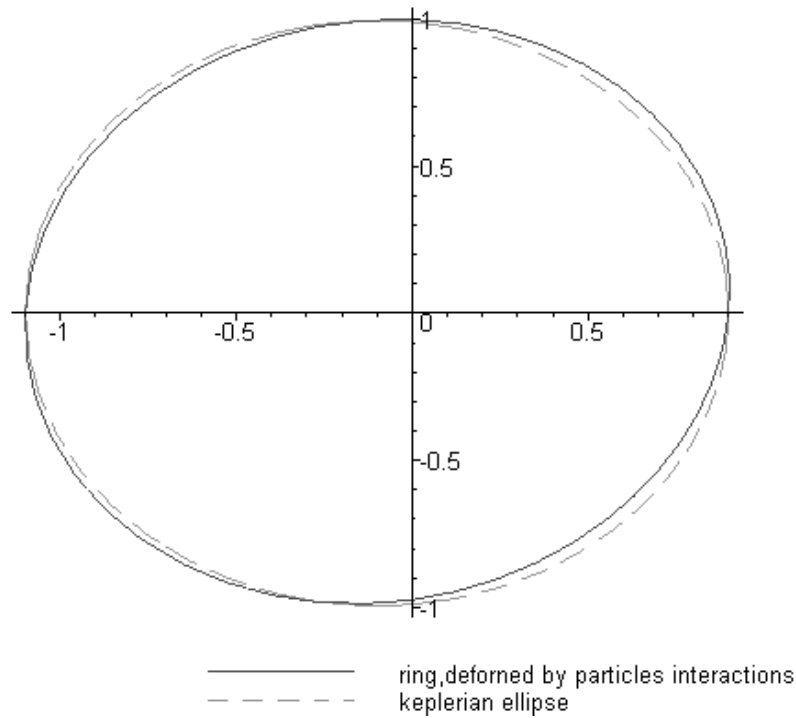

——— ring, deformed by particles interactions
– – – – keplerian ellipse

Fig.2. Eccentric ring shape change due to gravitational interactions of particles.

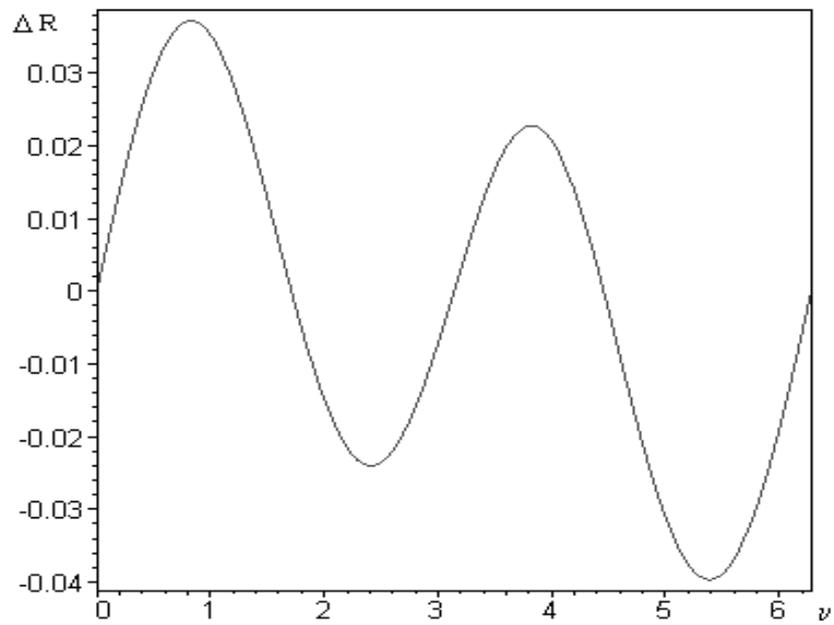

Fig.3. Declination from keplerian ellipse as function of true anomaly.

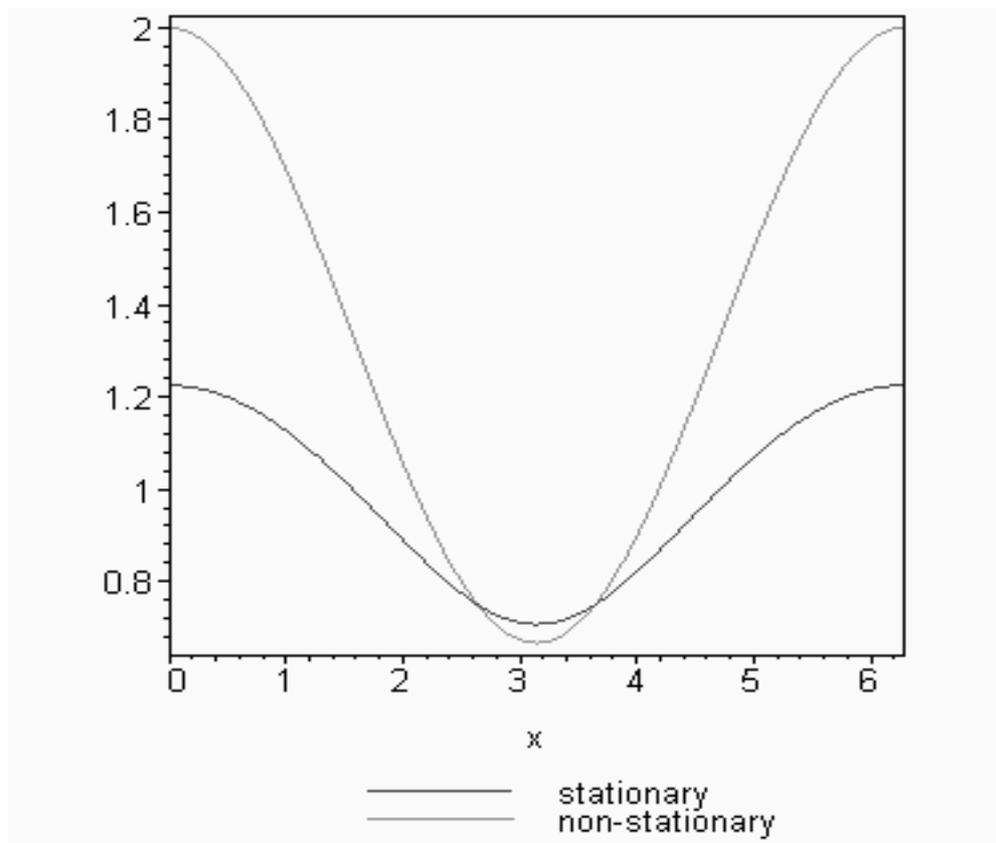

Fig.4. The stationary and non-stationary distribution of particles in eccentric ring. Dependence from true anomaly(x).

# RING PARAMETIC RESONANCE STRUCTURE. THEORETICAL MODEL

The equations of planar restricted Hill's problem may be reduced to a Hill equation for normal distance from variation orbit $x$ (Szebehely, 1967):

$$\frac{d^2x}{dt^2} + \omega^2(t)x = f(t) \qquad (6)$$

where $f(t)$ – known function of time, $\omega(t)$ - periodic function of time (to be determined).

It means, that area of parametric instability will appear in problem. Note, that variation orbit may be chosen by different way. In particular, it may be two-body problem solution.

Then, we enter gravitation interaction between particles, by modeling ring as a system of massive points. We shall consider ring consist of $N$ attractive particles, placed close to vertex of regular polygon. Each particle may slightly move around stationary position, and all system rotated with common angular velocity. Set the perturbing body in circular orbit with radius $r$ in ring plane. Let the radial perturbation of ring small $x \ll R$ where $R$ - radius of ring.

We show, that by taking into account interaction between ring's particles, it is possible to significantly change resonant structure in ring.

For the stability of motion investigation, an averaged equation may be used. An averaged angular moment is conserved. It means, that variations in mean motion are small, so the problem setting is restricted by conditions:

$$\delta\lambda \equiv \lambda - \lambda_s \approx \delta\varpi t = (\Omega - \omega_s)t \qquad (7)$$

where $\lambda$ and $\lambda_s$ - mean longitudes, $\Omega$ and $\omega_s$ - mean motions of perturbed and perturbing bodies accordingly, $t$ – time.

As it is evident, $\omega$ in (1) depend from time:

$$\omega^2 = \omega_0^2\left(1 + \sum_n h_n \cos n\delta\varpi\right) \qquad (8)$$

where $h_n$ – known coefficient. It is sufficient to consider Mathieu equations for the main mean motion resonance cases. At case circular perturbing body orbit:

$$\omega_0^2 = F(R) + \frac{d^2 U_{ring}}{dR^2} \qquad (9)$$

where $F(R)$ – known function of ring central distance $R$, $U_{ring}$ – ring perturbation function. The condition of appearance of parametric resonance and related instable zones:

$$\frac{2\omega_0}{(\Omega - \omega_s)} = n, \quad n = 1,2,3,\ldots \qquad (10)$$

where $\Omega$ and $\omega_s$ - mean motions of perturbed and perturbing bodies accordingly. Here $n$ – is an order of resonance.

So, in real planetary ring, perturbed by outer satellite, may appear parametric type of instability. The position of resonances is different than in case neglected particles gravity, because second term in (5) appears by ring's particles interaction. However, there is one very important effect: the center of instability is shifted relative exact commensurability.

## 3. RING STRUCTURE EXPLANATION

The expansion of rings perturbation has a form (Rosaev A.E. 2003):

$$U_r \equiv \frac{dU_{ring}}{dR} = -\sum_j^{n-1} \frac{\gamma m_j}{R^2(2\sin(\psi_j/2)^2} + \left(\sum_{j=1}^{n-1} \frac{3}{4}\frac{\gamma m_j}{R^3\sin(\psi_j/2)} - \frac{\gamma m_j}{R^3((2\sin(\psi_j/2))^3}\right)x + $$
$$\left(\sum_{j=1}^{n-1} \frac{3.75\gamma m_j}{R^4((2\sin(\psi_j/2))^3} - \frac{15}{4}\frac{\gamma m_j}{R^4((8\sin(\psi_j/2))}\right)x^2 + O(x^3) \tag{11}$$

where $\gamma$ - gravity constant, $m_j$ – mass of ring particle, $\psi_j = \frac{2\pi j}{N} + \phi_0$ - angular distance between $j$-particle and tested particle, $\phi_0$ - small arbitrary angle.

The dependence of perturbation function from distance relative ring's axis is shown at the Fig.5.

In general case, taking into account approximate expression for mean motions:

$$\Omega^2 \approx \frac{\gamma M}{R^3}, \quad \omega_s^2 \approx \frac{\gamma M}{r^3}, \tag{12}$$

where $M$ – central mass $R$ – ring radius, $r$ – perturbing satellite orbital radius.

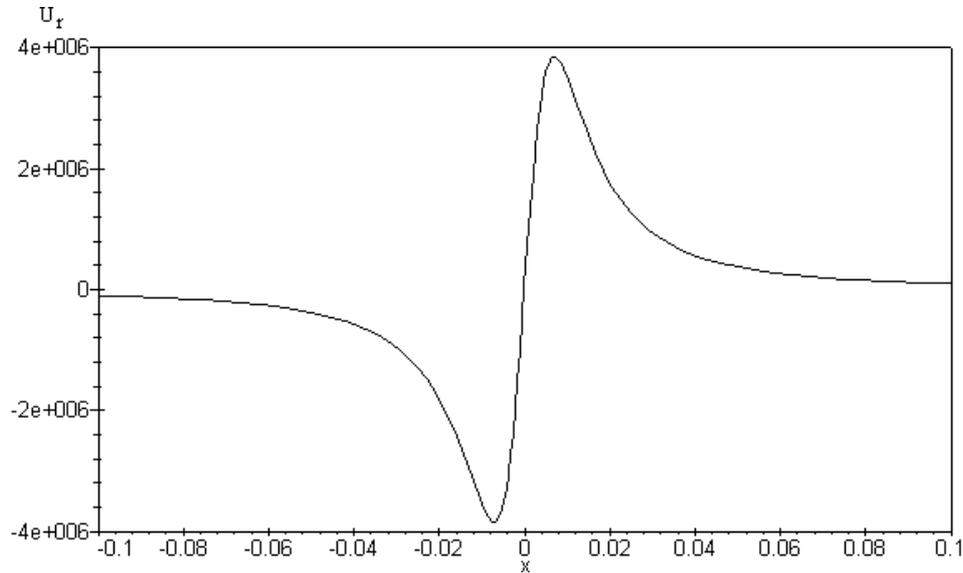

Fig.5. Ring radial perturbation function in dependence of distance from ring

The resonant condition for $\omega$ (10) may be rewritten in form:

$$\left(4\alpha - \frac{n^2 M}{r^3}\right)\beta^2 + \frac{2n^2 M}{r^{3/2}}\beta + 4M - n^2 M = 0 \quad \beta = R^{3/2} \tag{13}$$

Here α - ring particles gravity perturbation. It allow to calculate centers of instabilities positions:

$$R^{3/2} = r^{3/2} \frac{n^2 - 2\sqrt{n^2 + (n^2-4)\alpha r^3/M}}{n^2 - 4\alpha r^3/M} \qquad (14)$$

There are two cases at strong interaction between particles. The sign of α is determined by expansion (11). In case positive α we have instability relative small oscillations – a very narrow gap, shifted toward the planet. In case negative α we have oscillation with large amplitude: it is sufficient, if particle will be out first linear zone to remove it far from it's initial position. So, we have a broad area of instability, shifted away from central planet relative exact commensurability. **In result, we have three unstable gaps at each resonance.**

The coordinate of boundary point between positive and negative α is:

$$x_0 \approx R\left(-2\left(\frac{\pi}{N}\right)^2 \pm \frac{2}{\sqrt{5}}\frac{\pi}{N}\sqrt{1-\frac{\pi}{N}}\right) \qquad (15)$$

In case positive α it easy to be calculated:

$$\sum_{i=1}^{N-1} \frac{\gamma m_i}{(2R\sin(\pi i/N))^3} \approx 0.300512625 \frac{\gamma m_i N^3}{\pi^3 R^3} = 1.258784\gamma\rho\sigma^3 = \gamma\alpha \qquad (16)$$

where $\sigma = r_k N/(\pi R)$ - where $\sigma, \rho$ and $r_k$ - ring's «coefficient of filling», density and size of ring's particles accordingly, $N$ – number of particles. At $\sigma = 1$, for example, it is according with model maximal density, where particles in ring touch one another. As it seems from Fig.1, the negative α is about 0.5α positive.

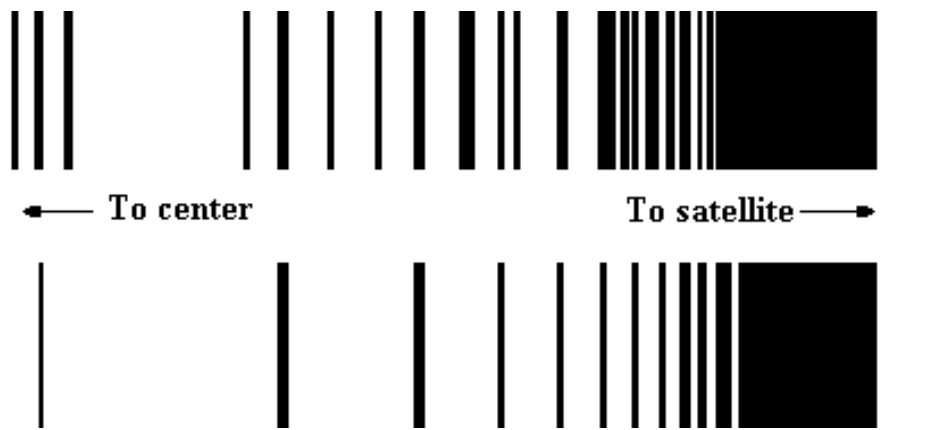

Fig. 6. Resonance structure at α=0.5

The effective excitation of parametric instability is possible since finite distance from planet (table 1).

Table 1.
Possible conditions of parametric instability in planetary rings

| Ring system | Range of distance ($10^3$ km) | Min distance for parametric instability ($10^3$ km) | Filling coefficient, $\sigma$ |
|---|---|---|---|
| Jupiter | 127-130 | 103 | 0.80 |
| Saturn | 70-140 | 69 | 0.49-0.98 |
| Uran | 41-51 | 36.5 | 0.72-0.90 |
| Neptun | 41-62 | 38.2 | 0.63-0.95 |

Density waves give the ability to estimate surface density ($\rho_{surf} \sim$ 60 g of material per square centimeter of ring area (Stone *et al.*, 1982.)) and filling coefficient. By using for Saturn mass $M=5.6846*10^{26}$ kg, $r_k$= 1 m, $\rho$ = 1 g/cm$^3$, the estimations of number of particles $N$, $\sigma$ and $\alpha$ gives:

$$N = \frac{3R\rho_{surf}}{2\rho r_k^2} \approx 9*10^7 \qquad \sigma = \frac{\rho_{surf}}{4\rho r_k} = 0.15 \div 0.3$$

$$\alpha = 1.25\rho\sigma^3 = 1.25\rho\left(\frac{Nr_k}{2\pi R}\right)^3 \approx 4 \div 12 \approx 0.01 \div 0.04 M/R^3 \qquad (17)$$

It is very important, that this condition not depend from ring (or ring's particle) mass, only from degree of filling ring by particles. ***It means, that even for low-massive rings, mutual gravitation interaction between particles can play a significant role in ring particles dynamics at small area close to ring.***

It is possible, now, to compare our model with observations. The shift of resonances due to particles gravity in dependence of particle size is given at Fig.7.

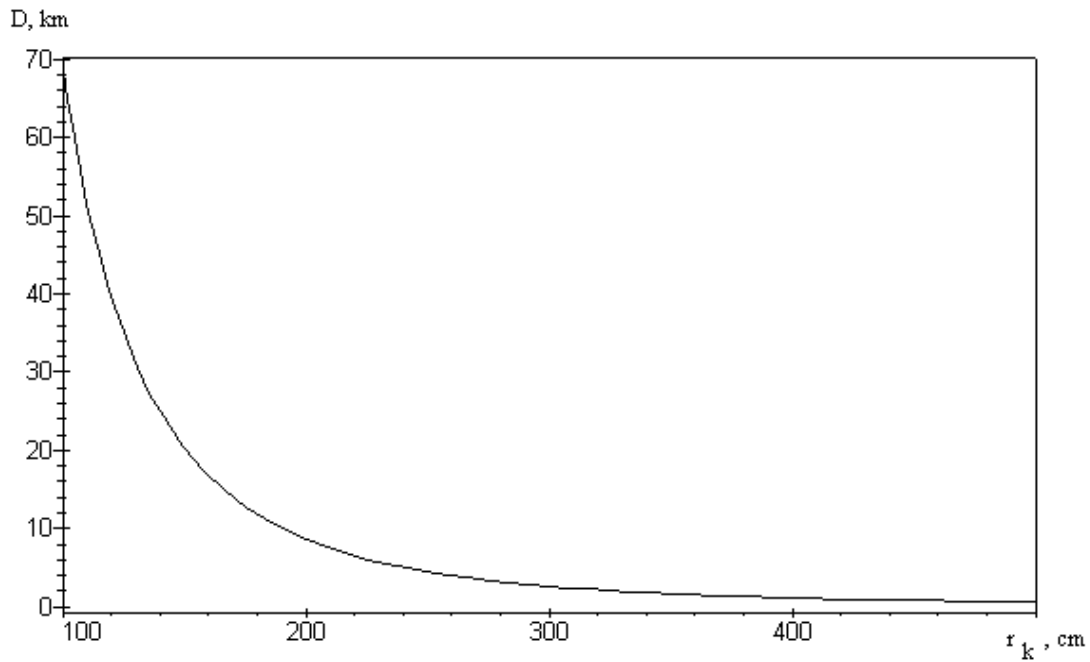

Fig 7. The shift of resonances due to particles gravity in dependence of particle size

It seems, that at observed Saturn system parameters, described model provide significant improvement of ring structure with Mimas mean motion resonances correlations. However, for better agreement, size of particles must be slightly smaller, about 0.1 – 0.5 meters. In this case, not only Cassini division, but Encke division (and maybe B ring inner edge) will coincide with resonance position.

## 4. CONCLUSIONS

The resonance perturbation of planetary ring by distant satellite is considered. The shift between simple mean motion resonances and parametric resonance zones is detected. This shift depends from ring properties. Unexpectedly, the resonance's structure depends only from particle's density, but not depends from particles (or ring) mass. It is shown, that mutual interaction between particles of ring occur significant effect on resonance structure. The results applied to Saturn ring system. By varying ring's particles parameters, is possible to explain observed shift between actual ring distances and exact commensurability.

It is predicted, that averaged size of ring particles (about 0.1-0.5 meters) slightly smaller, than accepted value 2-10 meters (Stone *et al.*, 1982).

The gravitation interaction between particles in rings can change shape of non-circular rings. The model of central configuration gives a way of estimation of this effect in dependence of ring particles properties. In case of massive particles, the stationary distribution of it along eccentric ring is differ from keplerian.